\documentstyle[12pt]{article}
\newcommand{\beq}{\begin{equation}}
\newcommand{\eq}{\end{equation}}
\newcommand{\bega}{\begin{eqnarray}}
\newcommand{\ega}{\end{eqnarray}}
\begin{document}
\title{Shell model on a random gaussian basis}
\author{K. Varga
\\
Institute  of Nuclear Research of the Hungarian Academy of
Sciences
\\
(MTA ATOMKI) Debrecen, Hungary
\\
and
\\
R.J. Liotta
\\
Royal Institute of Technology, S-10405 Stockholm, Sweden}
\maketitle
\date{\today}
\maketitle
\begin{abstract}
Pauli-projected random gaussians are used as a representation to solve
the shell model equations. The elements of the representation are
chosen by a variational procedure. This scheme is particularly suited
to describe cluster formation and cluster decay in nuclei. It
overcomes the basis-size problem of the ordinary shell
model and the technical difficulties of the cluster-configuration
shell model. The model reproduces the $\alpha$-decay width of $^{212}$Po
satisfactorily.
\end{abstract}

\vskip 48pt
\section {Introduction}
By complementing the shell model with a cluster-model basis,
the $\alpha$-decay of $^{212}$Po can be described very well \cite{Varga}.
This cluster-configuration shell model or hybrid model is, however, difficult
to apply. We have therefore developed a simpler model which is closer to
the conventional shell model but in practical applications
may prove to be far superior. The objective of this paper is
to report on this new model and its performance in the calculation of
the ground state of $^{212}$Po. To
this end, we compare its results with those of the hybrid model
\cite{Varga}.
\par\indent
The new model is essentially a shell model, in which the valence
orbits are replaced by combinations of gaussian
functions of different size parameters, with exact Pauli projection
off the core orbits. These gaussians are more flexible
than the harmonic oscillator single-particle (s.p.) functions because
their width parameters are free. A variational backgroud can guarantee
that they will be  automatically chosen
so as to suit to the problem considered. In particular, the gaussians
can produce a more realistic fall-off in the surface, which is important
for the description of decay processes.
Moreover, in the interior the gaussians are able to simulate the effects of
several oscillator shells. Note that the Pauli projection provides the
gaussians with some, say $n_0$, nodes, so that $n$ gaussians in the expansion
of a s.p. function will
involve s.p. orbits of $n_0,\ldots,n_0+n-1$ nodes. In spite of the inclusion of
such high-lying shells, the problem may still remain tractable if the
basis elements are carefully selected. We admit states in the basis depending
on their contribution to the g.s. energy. This method of choosing
random states as candidates for the basis is called \cite{VSL}
``stochastic variational method".
\par\indent

\section{The formalism}
The basic building block of our s.p. basis is the
gaussian (or nodeless harmonic oscillator function)
of size parameter $\nu$, defined by
\beq
\varphi_l^\nu (r)=\left(
{2^{l+2} (2\nu)^{l+3/2}
\over \sqrt{\pi} (2l+1)!!}\right)^{1/2}
r^l e^{-\nu r^2} .
\eq
As any well-behaved function can be approximated with
any prescribed precision by a linear combination of gaussians of different size
parameters, $\sum_{i} c_i \varphi_l^{\nu_i} (r)$, one can use
such combinations to approximate the valence s.p. functions of the shell-model
Hamiltonian.
\par\indent
These gaussians are not orthogonal to the core orbits.
To take care of antisymmetrization, one can introduce
the operator that projects out the s.p. states occupied in the core
\cite{Varga,Horiuchi}:
\beq
P_i=1-\sum_{j=5}^{A}\vert \phi^j(i)\rangle \langle
\phi^j(i)\vert
\eq
The index $i$ distinguishes the proton ($i=1,2$)
and neutron ($i=3,4$) orbits.
\par\indent
The Pauli-correct s.p. wave functions are then constructed as
\beq
{\hat \varphi_{lj}^\nu} (i)=
P_i\left\{\varphi_l^\nu (r_i) \left[ Y_{l} ({\hat r}_i)
 \chi_{1 \over 2} (i) \right]_{jm}\right\}
\label{eq:trial1}
\eq
where $\chi_{{1 \over 2}\sigma}(i)$ is the spin function of
the $i$th nucleon.
\par\indent
In the description of $^{212}$Po it is reasonable to assume the core,
$^{208}$Pb, to be passive and massive. (This approximation
could be removed by using the formalism of the cluster-orbital shell
model \cite{Suzuki}.)
\par\indent
As these s.p. functions are fully analogous to
those in the conventional shell model,
the many-particle basis can be constructed in the usual
way. In proton-neutron formalism and $jj$ coupling scheme
the two-proton and two-neutron functions are given by
\beq
\Psi_{\pi}^{i}={\cal A}\left[
{\hat \varphi_{l_1 j_1}^{\nu^i_1}}(1){\hat \varphi_{l_2 j_2}^{\nu^i_2}}(2)
\right]_{J_\pi}
\label{eq:trial21}
\eq
and
\beq
\Psi_{\nu}^{j}={\cal A}\left[
{\hat \varphi_{l_3 j_3}^{\nu^j_3}}(3){\hat \varphi_{l_4 j_4}^{\nu^j_4}}(4)
\right]_{J_\nu} ,
\label{eq:trial22}
\eq
where the operator ${\cal A}$ antisymmetrizes between
like particles, $\pi=\{l_1,j_1,l_2,j_2,J_\pi\}$,
$\nu=\{l_3,j_3,l_4,j_4,J_\nu\}$. A four-particle state is formed as
\beq
\Psi_{(\pi\nu)JM}^{k}=\left[\Psi_{\pi}^{i}\Psi_{\nu}^{j}\right]_{JM}.
\label{eq:trial4}
\eq
The trial function is then
\beq
\Psi_{JM}=\sum_{\pi\nu k} c_k \Psi_{(\pi\nu)JM}^{k}.
\eq
\par\indent
To describe the dynamics of the valence particles, we minimize
the expectation value of their Hamiltonian
\beq
H=\sum_{i=1}^{4} (T_i+U_i)+\sum_{1\le i < j \le 4} V_{ij},
\eq
where $T$ is the kinetic energy, $U$ is the core-particle interaction
and $V$ is the particle-particle interaction.
\par\indent
The trial function $\Psi_{JM}$ contains a great number of terms
belonging to different configurations and size parameters.
Since the dependence of the eigenvalue problem upon the
variational parameters $c_i$ is linear one can determine them by
finding the lowest eigenvalue of a generalized linear algebraic
eigenvalue problem, i. e.
\beq
{\bf H} {\bf c} = E_{N} {\bf A} {\bf c},
\label{eq:ei}
\eq
where ${\bf H}$ and ${\bf A}$ are the Hamiltonian and overlap matrices.
The subscript $N$ stands for the dimension of the eigenvalue problem.
The overlap matrix ${\bf A}$ is nondiagonal
due to the nonorthogonality of the functions $\Psi_{(\pi\nu)JM}^{k}$.
\par\indent
This variational approach is only expected to yield a good approximation to
the g.s. energy if the size
parameters $\nu_1^k,...,\nu_4^k$ are adequately chosen.
Previous experience with the stochastic multicluster model \cite{VSL}
shows that the number of basis states needed is not excessive
even in the five-cluster case, which is comparable with the present
core plus four-nucleon system.

We generated trial basis states
randomly, and singled out, by an admittance test, the ones whose contribution
to the g.s. energy was larger than a preset limit, $\varepsilon$ (``utility
testing" \cite{VSL}).
That is, we start (k=1) by making a random choice of the size
parameter quartet $\nu_1^k,...,\nu_4^k$ from the physically important
$\nu$ region, and set up a basis state of a particular configuration.
The energy expectation value gives the first approximate
energy, $E_1$. Then we generate a new random set $\nu_1^2,...,\nu_4^2$,
and solve eq. (\ref{eq:ei}) for $N=2$.
Due to the variational character of the method any parameter quartet
improves the energy ($E_2 < E_1$), but we only adopt
the basis element if $E_1-E_2>\varepsilon$.
This procedure is then repeated with new random elements
up to convergence, and then the basis is enlarged by including further
configurations.

At each trial step the energy can be determined
by solving the eigenvalue problem using any conventional numerical
algorithm. But as we are only interested in the g.s. energy,
and we proceed stepwise,
we can use the much more economical modified Jacobi method \cite{raedt}.
The Jacobi method (see e. g. Ref. \cite{jacobi}) solves the eigenvalue
problem ${\bf H}{\bf C}= E {\bf C}$ eliminating the nondiagonal elements
by successive plane rotations.
In the modified Jacobi method, the lowest eigenvalue is only
determined by elimination of the nondiagonal elements corresponding to
that eigenvalue \cite{raedt}. Thus it uses, in every step, the results of the
diagonalization of the previous step.
Although the modified Jacobi method can be generalized to solve fully the
eigenvalue problem \cite{raedt}, we found it
more convenient to re-orthogonalize our basis in each step of the random search
procedure.
This is advantageous because $\bf A$ is block-diagonal (each
configuration provides a block) and we need to orthogonalize only
within the configuration whose basis is being enlarged.
Since the number of basis states in any configuration is low, this
re-orthogonalization requires almost no computational effort.
\par\indent
A delicate point of this procedure is
the choice of $\varepsilon$ because a
too large value bogs down the procedure,
whereas a too small value results in highly redundant bases of large
dimensions. It is expedient \cite{raedt} to re-adapt the acceptance criterion
dynamically. For instance, in the present calculation we started with
$\varepsilon=0.01$ MeV, and after 10 failed attempts to find a good enough
basis state, we reduced $\varepsilon$ by 2, and this was repeated
two more times.
In this way we managed to find further states of some significance.
This procedure has been thorougly tested \cite{VSL}, and it was
found that it is extremely powerful and reliable \cite{VSL}.

\section{Details}

The s.p. potentials include scalar and spin-orbit
Woods-Saxon parts as well as the Coulomb interaction.
To facilitate the analytical calculation of the matrix elements, all
potential terms are expanded in terms of gaussians.
To be able to compare the
``pure shell model''\cite{Varga} with the present model, we use the
potential parameter set B of ref. \cite{Varga}
and employ the same Pauli projector as there, that is the occupied s.p.
states are represented by oscillator functions of size
parameter $\nu_0=0.083 {\rm fm}^{-2}$.
\par\indent
For the valence orbits we selected the size parameters from the physically
important interval $4\ {\rm fm} \le \nu^{-1/2} \le 14\ {\rm fm}$.
The following s.p. quantum numbers were used:
\bega
l,j & = & h_{9/2}, f_{7/2}, i_{13/2}, f_{5/2}, p_{3/2}  \ \ (protons)
\nonumber \\
    & = & g_{9/2}, i_{11/2}, j_{15/2}, d_{5/2}, s_{1/2}, g_{7/2}
         \ \  (neutrons) \nonumber.
\ega
Just as in the former model, we set up the four-particle basis by coupling all
possible two-particle combinations up to $J_{\pi}=8$ and $J_{\nu}=8$.
\par\indent
We build up the basis by the method skecthed above
in each configuration $\{\pi \nu \}$.
We ordered the configurations according to their
expected contributions to the energy, i.e. according to the
sums of the s.p. energies involved and, among states involving
the same s.p. states, to the
two-particle angular momenta in increasing order.
Thus we begin with the configurations
$[[h_{9/2}]^2_{J_{\pi}} [g_{9/2}]^2_{J_{\nu}}]_0$,
with $J_{\pi}=J_{\nu}=0,1,2,\ldots$,
$[[h_{9/2}]^2_{J_{\pi}} [i_{11/2}]^2_{J_{\nu}}]_0$,
with $J_{\pi}=J_{\nu}=0,1,2,\ldots$,
$[[f_{7/2}]^2_{J_{\pi}} [g_{9/2}]^2_{J_{\nu}}]_0$,
with $J_{\pi}=J_{\nu}=0,1,2,\ldots$ etc.
If a configuration cannot contribute to the g.s. energy by at least
$\varepsilon/2^3$, then it is omitted.
\par\indent
\section{Results}

First we solved the eigenvalue problem of
the s.p. part of the Hamiltonian on the basis (\ref{eq:trial1}),
The scalar depth
was adjusted so as to best reproduce the experimental s.p. energies,
as done in ref. \cite{Varga}.
We obtained this with the scalar strengths $V_p=61.60$ MeV
and $V_n=43.80$ MeV for protons and neutrons,
respectively. The resulting s.p. energies are in excellent
agreement with the ones obtained by a direct numerical integration.
\par\indent
We then calculated the g.s. ($J=0$) energy of $^{212}$Po. We
terminated the random search at a basis dimension $N=397$ and a
number of configurations $n=118$. As the number of basis states is
increased the g.s. energy
converges to $E=-19.25$ MeV (the experimental value is
$-19.35$ MeV.), as seen in figure 1.
This may be compared  with the value
$E=-18.96$ MeV of the pure shell model or
the value $E=-19.18$ MeV in the hybrid model \cite{Varga}.

To check the convergence we repeated the calculation
using different random parameter sets with dimensions
$N=413$, $n=120$ and $N=380$, $n=127$. The three energies agree within
0.04 MeV.
To see the role of the order of the configurations, we repeated again
the calculations with reversed order. The
resulting energies are very close to the previous ones, i. e.
$-19.24$ MeV, $-19.28$ MeV and $-19.25$ MeV,
but the basis sizes are almost doubled.
To test the choice of $\varepsilon$, we made three calculations with
$\varepsilon=0.005$ MeV. The lowest calculated energy obtained is $E=-19.27$
MeV, in good agreement with the former calculations, but the
basis sizes increased considerably ( $N=631$, $n=193$).

\par\indent
We also calculated the spectroscopic (or formation) amplitude defined by
\beq
g(r)= \langle \Phi^{(r)} \vert \Psi_{00} \rangle,
\eq
with
\beq
\Phi^{(r)}=\Phi_{\alpha} {\delta(r-r_{c\alpha}) \over r_{c\alpha}}
Y_{LM}({\hat r}_{c\alpha}),
\eq
where $\Phi_{\alpha}$ is the intrinsic wave function of the $\alpha$ particle
constructed from a $0s$ harmonic-oscillator Slater determinant with size
parameter $\nu_\alpha$. (See Ref. \cite{Varga} for details.)
The amplitudes of the present model and of the cluster-configuration shell
model are compared in Fig. 2. The main peak is smaller and its position
is somewhat closer to the origin (at 8.0 fm) in the present model
than in the cluster-configuration model (8.2 fm).

The present spectroscopic factor is 0.011, while it is 0.023 in the
hybrid model.
\par\indent
To calculate the decay width $\Gamma$, we use the
$R$-matrix formula \cite{width}
\beq
\Gamma=2 P(a) {\hbar^2 \over 2 M_{\alpha} a} g^2(a),
\eq
where $M_{\alpha}$ is the reduced mass, $a$ is the channel
radius and $P$ is the Coulomb penetration factor.
The width, calculated in the interval $10  {\rm fm} \le a \le 12$ fm,
where the effect of the nuclear forces and Pauli exchanges are negligible,
is $\Gamma=7.2\times 10^{-16}$ MeV. This value is about one
third of the hybrid-model value ($\Gamma=2.1\times 10^{-15}$)
and about one half of the experimental ($\Gamma=1.5\times 10^{-15}$) value.
Note that the conventional shell model, like that of ref. \cite{Varga},
give much smaller and spacially confined amplitudes, while
the shell models with sophisticated truncation schemes
\cite{Arima,Gordana} give amplitudes of comparable extensions but smaller
spectroscopic factors and decay widths.

\section{Summary}
With the help of a stochastic variational method, a flexible shell-model wave
function has been constructed for the g.s. of $^{212}$Po using
combinations of gaussian functions as
s.p. orbits. The present model gives nearly as
good a result for the $\alpha$-decay width
as our former cluster-configuration shell model, which is sharpened just to
describe $\alpha$-decay. The chief merit of the model presented here is its
technical simplicity, which makes it easily applicable
to other $\alpha$-decays or even to heavy-cluster decays.

This work was supported by the OTKA grants (No. 3010 and F4348). We are
grateful to Prof. R. G. Lovas for useful discussions.

\newpage\eject
\noindent
{\Large \bf Figure captions}
\vskip 48pt
\par\noindent
{\bf Figure 1}
Convergence of the energy of $^{212}$Po (g.s.) as a function
of the basis dimension $N$.
\par\noindent
{\bf Figure 2}
Formation amplitude g(r) (eq. (10)) as a function of the
distance between the decaying $\alpha$-particle and the daughter
nucleus $^{208}$Pb (g.s.)
\end{document}